\documentclass[dvips]{acta}
\usepackage{supertabular,lscape,epsfig}
\usepackage{amssymb}
\usepackage{amsmath}
\DeclareSymbolFont{ppa}{OT1}{ppl}{m}{it}
\DeclareMathSymbol{\vv}{\mathalpha}{ppa}{'166}

\newfont{\hb}{rphvb at 10pt}%bezszeryfowe półgrube
\newfont{\hbo}{rphvbo at 10pt}%bezszeryfowe półgrube kursywa
\newfont{\bitt}{rptmbi at 12pt}%półgruba kursywa (tytuł artykułu)
\newfont{\bits}{rptmbi at 11pt}%półgruba kursywa (tytuły rozdziałów)

\SetPages{429}{451}

\SetVol{63}{2013}

\begin{document}

%Zwarte naglowki, jeden wiersz, appendix
\newcommand{\TabApp}[2]{\begin{center}\parbox[t]{#1}{\centerline{
  {\bf Appendix}}
  \vskip2mm
  \centerline{\small {\spaceskip 2pt plus 1pt minus 1pt T a b l e}
  \refstepcounter{table}\thetable}
  \vskip2mm
  \centerline{\footnotesize #2}}
  \vskip3mm
\end{center}}

%Zwarte naglowki, jeden wiersz
\newcommand{\TabCapp}[2]{\begin{center}\parbox[t]{#1}{\centerline{
  \small {\spaceskip 2pt plus 1pt minus 1pt T a b l e}
  \refstepcounter{table}\thetable}
  \vskip2mm
  \centerline{\footnotesize #2}}
  \vskip3mm
\end{center}}

%Zwarte naglowki, dwa wiersze
\newcommand{\TTabCap}[3]{\begin{center}\parbox[t]{#1}{\centerline{
  \small {\spaceskip 2pt plus 1pt minus 1pt T a b l e}
  \refstepcounter{table}\thetable}
  \vskip2mm
  \centerline{\footnotesize #2}
  \centerline{\footnotesize #3}}
  \vskip1mm
\end{center}}

%Zwarte naglowki, jeden wiersz, appendix
\newcommand{\MakeTableApp}[4]{\begin{table}[p]\TabApp{#2}{#3}
  \begin{center} \TableFont \begin{tabular}{#1} #4 
  \end{tabular}\end{center}\end{table}}

%Zwarte naglowki, jeden wiersz
\newcommand{\MakeTableSepp}[4]{\begin{table}[p]\TabCapp{#2}{#3}
  \begin{center} \TableFont \begin{tabular}{#1} #4 
  \end{tabular}\end{center}\end{table}}

%Zwarte naglowki, jeden wiersz
\newcommand{\MakeTableee}[4]{\begin{table}[htb]\TabCapp{#2}{#3}
  \begin{center} \TableFont \begin{tabular}{#1} #4
  \end{tabular}\end{center}\end{table}}

%Zwarte naglowki, dwa wiersze
\newcommand{\MakeTablee}[5]{\begin{table}[htb]\TTabCap{#2}{#3}{#4}
  \begin{center} \TableFont \begin{tabular}{#1} #5 
  \end{tabular}\end{center}\end{table}}

%{\it Acta Astronomica Archive}
%\parskip=0pt \itemsep=1mm \setlength{\itemsep}{0.4mm}\setlength{\parindent}{-1em} \setlength{\itemindent}{-1em} - po \begin{itemize} - wszystko
%FWHM, PSF, S/N - proste, 
%MgII, H$\alpha$
%rms, rhs, sd - kursywa
%{\sc DAOPhot}
%{\sc Fnpeaks}
%{\sf files}
%Galactic wszystko (bulge, center, plane, disk, coordinates, latitudes...)
%Cepheids
%type~ Cepheids, Population~II Cepheids
%a.u.
\newfont{\bb}{ptmbi8t at 12pt}
\newfont{\bbb}{cmbxti10}
\newfont{\bbbb}{cmbxti10 at 9pt}
\newcommand{\uprule}{\rule{0pt}{2.5ex}}
\newcommand{\douprule}{\rule[-2ex]{0pt}{4.5ex}}
\newcommand{\dorule}{\rule[-2ex]{0pt}{2ex}}
\def\thefootnote{\fnsymbol{footnote}}
\begin{Titlepage}
\Title{Variable Stars in the Globular Cluster NGC~288: [Fe/H] and
Distance\footnote{ Based on observations collected with the 2.0~m telescope
at the Indian Astrophysical Observatory, Hanle, India.}}
\Author{A.~~A~r~e~l~l~a~n~o~~F~e~r~r~o$^{1,2}$,~~ 
D.\,M.~~B~r~a~m~i~c~h$^{2,3}$,~~ 
S.~~G~i~r~i~d~h~a~r$^4$, ~~
R.~~F~i~g~u~e~r~a~~J~a~i~m~e~s$^{2,5}$,~~ 
N.~~K~a~i~n~s$^2$~~ and~~
K.~~K~u~p~p~u~s~w~a~m~y$^4$}
{$^1$Instituto de Astronom\'ia, Universidad Nacional Aut\'onoma de M\'exico\\
e-mail:armando@astro.unam.mx\\
$^2$European Southern Observatory, Karl-Schwarzschild-Stra{\ss}e 2, 85748~Garching~bei~M\"unchen, Germany\\
$^3$Qatar Environment and Energy Research Institute, Qatar Foundation, 
Tornado Tower, Floor~19, P.O. Box~5825, Doha, Qatar\\
$^4$Indian Institute of Astrophysics, Koramangala 560034, Bangalore, India\\
$^5$SUPA, School of Physics and Astronomy, University of St~Andrews, North Haugh, St~Andrews, KY16~9SS, United Kingdom}
\Received{November 19, 2013}
\end{Titlepage}

\Abstract{A search for variable stars in the globular cluster NGC~288 was
carried out using a time-series of CCD images in the {\it V} and {\it I}
filters. The photometry of all stellar sources in the field of view of our
images, down to $V\approx19$~mag, was performed using difference image
analysis (DIA). For stars of $\approx15$~mag, measurement accuracies of
$\approx8$~mmag and $\approx10$~mmag were achieved for {\it V} and {\it I}
respectively. Three independent search strategies were applied to the 5525
light curves but no new variables were found above the threshold limits
characteristic of our data set. The use of older data from the literature
combined with the present data allowed the refinement of the periods of all
known variables. Fourier light curve decomposition was performed for the
RRab and the RRc stars to obtain an estimate of ${\rm
[Fe/H]}_{ZW}=-1.62\pm0.02$ (statistical) $\pm0.14$ (systematic). A true
distance modulus of $14.768\pm0.003$~mag (statistical) $\pm0.042$~mag
(systematic), or a distance of $8.99\pm0.01$~kpc (statistical)
$\pm0.17$~kpc (systematic) was calculated from the RRab star. The RRc star
predicts a discrepant distance about one kiloparsec shorter but it is
possibly a Blazhko variable. An independent distance from the P--L
relationship for SX~Phe stars leads to a distance of $8.9\pm0.3$~kpc. The
SX~Phe stars V5 and V9 are found to be double mode pulsators.}{globular
clusters: individual: NGC~288 -- Stars: variables: RR~Lyrae, Stars: variables: delta~Scuti}

\Section{Introduction}
The globular cluster NGC~288 (C0050-268 in the IAU nomenclature) ($\alpha=
00\uph52\upm45\zdot\ups2$, $\delta=-26\arcd34\arcm57\zdot\arcs4$, J2000;
$l=151\zdot\arcd29$, $b=-89\zdot\arcd38$) lies toward the southern
Galactic pole at about 12~kpc from the Galactic center, hence it is subject
to very little interstellar reddening, $E(B-V)=0.03$~mag (Harris 1996, 2010
update).

The cluster has been studied photometrically since 1943, which has led to
the discovery of several variable stars. Using 144 photographic plates from
the Franklin-Adams camera at Johannesburg, Oosterhoff (1943) reported the
discovery of a long-period ($\approx100$~d) semi-regular variable star V1
as the first variable star in NGC~288. The next variable star V2, an RRab
star, was discovered about thirty-five years later by Hollingsworth and
Liller (1977) using seventeen {\it B} photographic plates and they
estimated a period of $\approx0.679$~d. It was not until the era of CCD
cameras that NGC~288 was studied again for variable stars in a pair of
papers by Janusz Kaluzny and collaborators (Kaluzny 1996, Kaluzny,
Krzeminski and Nalezyty 1997). These investigators used PSF-fitting
photometry to find a new RR~Lyr star V3 pulsating in the first-overtone,
six SX~Phe pulsators (V4--V9), and one eclipsing binary of the W~UMa type
(V10). Although Pietrukowicz \etal (2008) used difference image analysis
photometry to search (unsuccessfully) for dwarf novae in time-series CCD
images of NGC~288, they did not analyse their data for the known variables
or attempt to find any new variables.

Our study to search for variable stars in NGC~288 and our analysis of their
characteristics is therefore the first such study using CCD image data and
DIA combined.

The distribution of these ten variables in the cluster is rather peculiar as
they define an off-center concentration of less than 3~arcmin in diameter
in the otherwise $10\times10$~arcmin$^2$ field of our images. While this
distribution may not be very improbable, as will be discussed later in the
paper, it added to our interest in exploring the possibility of an
incomplete census of variables stars. In the recent past our team has
exploited the powerful technique of DIA for time-series CCD images in
globular clusters to update and characterise the population of variables in
them (\eg Arellano Ferro \etal 2013, 2011, Kains \etal 2013, 2012, Figuera
Jaimes \etal 2013, Bramich \etal 2011).

In the present paper we report the results of our variable star search in
the {\it V} and {\it I} filters and the calculation of the cluster
metallicity and distance. In Section~2 we describe the observations and
data reductions. In Section~3 we describe the approaches used to identify
new variables and the procedure to refine the periods. In Section~4 we
apply Fourier light curve decomposition to the RR~Lyr stars V2 and V3 and
calculate their metallicity and absolute magnitude, and hence the distance
to the cluster. In Section~5 we discuss the SX~Phe P--L (Period-Luminosity)
relation and use the SX~Phe stars as independent indicators of the cluster
distance. The double mode nature of V5 and V9 is also discussed there. In
Section~6 the properties of the variables V1 and V10 are addressed. In
Section~7 we summarise our results.

\Section{Observations and Reductions}
\label{sec:Observations}
The observations employed in the present work were performed using the
Joh\-nson--Kron--Cousins {\it V} and {\it I} filters on nine nights during
2010--2013 at the 2.0~m telescope of the Indian Astronomical Observatory
(IAO), Hanle, India, located at 4500~m above sea level. The detector was a
Thompson CCD of $2048\times2048$ pixels with a pixel scale of
0.296~arcsec/pix translating to a field of view (FoV) of
$\approx10.1\times10.1$~arcmin$^2$.

The log of observations is shown in Table~1 in which the dates, number of
frames, exposure times and average nightly seeing are recorded. A total of
174 epochs in the {\it V} filter and 201 in the {\it I} filter spanning
almost three years were obtained.

\MakeTableee{lrcrcc}{12.5cm}{The distribution of observations of NGC~288 for
each filter}
{\hline
\noalign{\vskip3pt}
Date [y.m.d]  &  $N_V$ & $t_V$ [s] & $N_I$ & $t_I$ [s] & Avg seeing [\arcs] \\
\noalign{\vskip3pt}
\hline
\noalign{\vskip3pt}
2010.12.11 & 21   & 30--80   & 36    &20--25 &1.8\\
2010.12.12 & 8    & 80       & 15    &20--60 &1.7\\
2011.10.07 & 20   & 80--100  & 19    &15--20 &2.5\\
2011.11.02 & 23   & 80--100  & 24    &15--20 &2.7\\
2011.11.03 & 33   & 80--150  & 33    &20--30 &2.5\\
2011.11.05 & 42   & 80       & 42    &20     &2.5\\
2013.01.20 & 2    & 160      & 2     &30     &2.7\\
2013.08.26 & 18   & 35-45    & 20    &10--25 &1.7\\
2013.08.27 & 7    & 70-90    & 10    &10--20 &1.9\\
\noalign{\vskip3pt}
\hline
\noalign{\vskip3pt}
Total:     &174   &          &201    &      &\\
\noalign{\vskip3pt}
\hline
\noalign{\vskip3pt}
\multicolumn{6}{p{7.9cm}}{Columns $N_V$ and $N_I$ represent the number of
images taken with the {\it V} and {\it I} filters respectively. Exposure
time, or range of exposure times, employed during each night for each
filter are listed in the columns $t_V$ and $t_I$ and the average seeing in
the last column.}
}

\subsection{Difference Image Analysis}
We employed the technique of difference image analysis (DIA) to extract
high-precision photometry for all of the point sources in the images of
NGC~288 and we used the {\sc DanDIA}\footnote{{\sc DanDIA} is built from
the {\sc DanIDL} library of IDL routines available at {\it
http://www.danidl.co.uk}} pipeline for the data reduction process
(Bramich \etal 2013) which includes an algorithm that models the
convolution kernel matching the PSF of a pair of images of the same field
as a discrete pixel array (Bramich 2008).

In general, a reference image is built by stacking a set of images with the
best seeing. However, in the present case the seeing was not particularly
good and we opted for using only one image as a reference image for each
filter. Then a sequence of difference images was created by subtracting the
relevant reference image, convolved with an appropriate spatially variable
kernel, from each registered image. The spatially variable convolution
kernel for each registered image was determined using bilinear
interpolation of a set of kernels that were derived for a uniform
$6\times6$ grid of subregions across the image.

The differential fluxes for each star detected in the reference image were
measured on each difference image.  Light curves for each star were
constructed by calculating the total flux $f_{\rm tot}(t)$ in ADU/s at each
epoch $t$ from:
$$f_{\rm tot}(t)=f_{\rm ref}+\frac{f_{\rm diff}(t)}{p(t)}\eqno(1)$$
where $f_{\rm ref}$ is the reference flux [ADU/s], $f_{\rm diff}(t)$ is
the differential flux [ADU/s] and $p(t)$ is the photometric scale factor
(the integral of the kernel solution).  Conversion to instrumental
magnitudes was achieved using:
$$m_{\rm ins}(t)=25.0-2.5\log\left[f_{\rm tot}(t)\right]\eqno(2)$$
where $m_{\rm ins}(t)$ is the instrumental magnitude of the star at time
$t$. Uncertainties were propagated in the correct analytical fashion.

The above procedure and its caveats have been described in detail in
Bramich \etal (2011).

We also performed a relative self-calibration of the ensemble
photometry. We applied the methodology developed in Bramich and Freudling
(2012) to solve for the magnitude offsets $Z_k$ that should be applied to
each photometric measurement from the image $k$. In terms of DIA, this
translates into a correction (to first order) for the systematic error
introduced into the photometry from an image due to an error in the fitted
value of the photometric scale factor $p$. We found that in either filter
the magnitude offsets that we derive are of the order of
$\approx1{-}10$~mmag with a handful of worse cases reaching
$\approx30$~mmag.  Applying these magnitude offsets to our DIA photometry
notably improves the light-curve quality, especially for the brighter
stars.

\subsection{Transformation to the Standard System}
Standard stars in the field of NGC~288 are very numerous in the online
collection of Stetson (2000)\footnote{\it
http://www3.cadc-ccda.hia-iha.nrc-cnrc.gc.ca/community/STETSON/standards}.
We selected a group of standards in the FoV of our images that cover the
{\it V} and $V-I$ ranges between 14.5~mag and 20.5~mag and 0~mag and
1.5~mag, respectively, to ensure a good transformation for most of the full
Color--Magnitude Diagram (CMD), and to check for the color dependence of
the transformations.

\begin{figure}[htb] 
\centerline{\includegraphics[width=9cm]{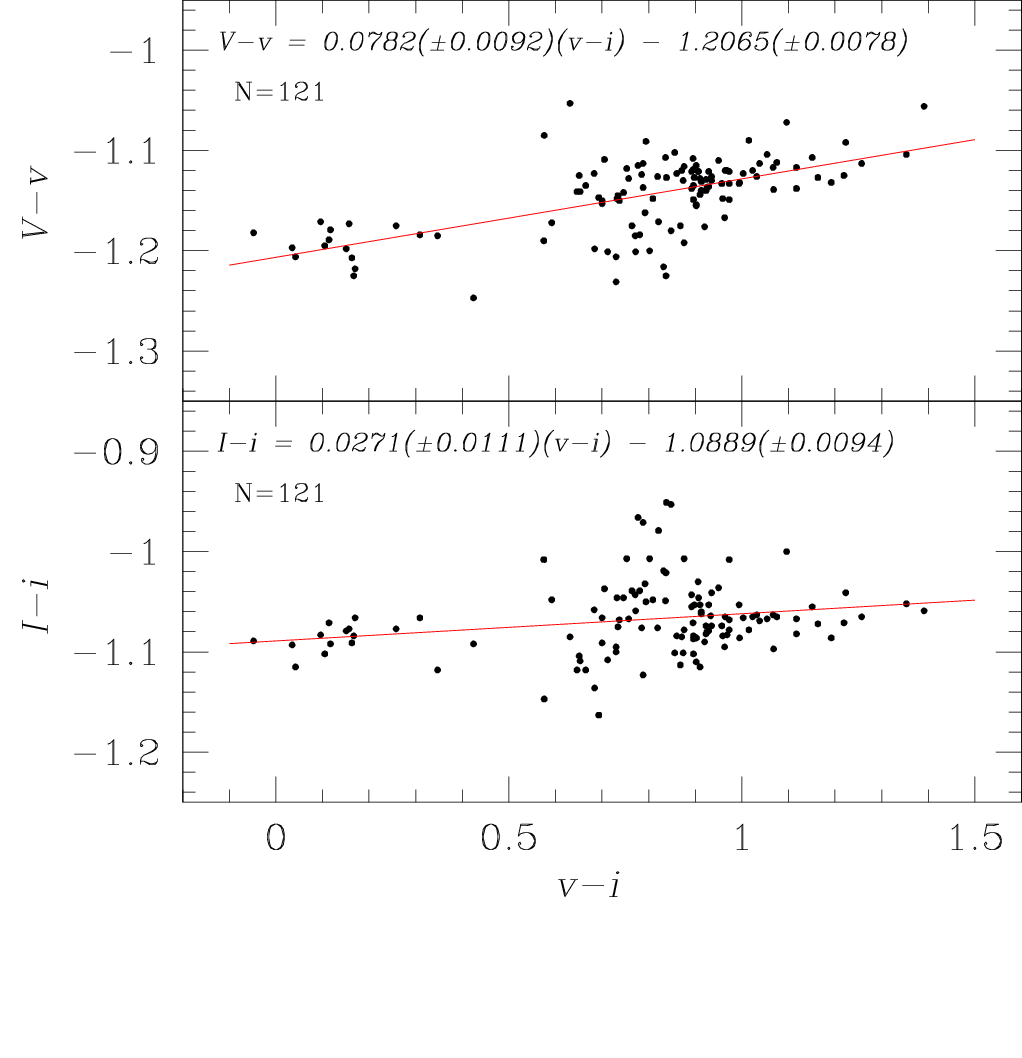}}
%\centerline{\includegraphics[width=1.0\textwidth,natwidth=8cm,natheight=8cm]{fig1.ps}}
\FigCap{Transformation relations between the instrumental and the standard
photometric systems using a set of standard stars in the field of NGC~288
from the collection of Peter Stetson. {\it Top} and {\it bottom panels}
correspond to the observations in the {\it V} and {\it I} filters
respectively. See Section~2.2 for a discussion.}
\end{figure}
\renewcommand{\TableFont}{\scriptsize}
\MakeTable{c@{\hspace{7pt}}c@{\hspace{7pt}}c@{\hspace{7pt}}c@{\hspace{7pt}}c@{\hspace{7pt}}c@{\hspace{7pt}}c@{\hspace{7pt}}c@{\hspace{7pt}}c@{\hspace{7pt}}c@{\hspace{7pt}}c}{12.5cm}{Time-series {\it V} and {\it I} photometry for all the confirmed variables in our field of view}
{\hline
\noalign{\vskip3pt}
Variable& Filter & HJD & $M_{\rm std}$ & $m_{\rm ins}$ & $\sigma_m$ & $f_{\rm ref}$   & $\sigma_{\rm ref}$ & $f_{\rm diff}$ & $\sigma_{\rm diff}$ & $p$ \\
Star ID 
&        
& [d] 
& {\tiny [mag]}
& {\tiny [mag]}
& {\tiny [mag]}
& {\tiny [ADU/s]} 
& {\tiny [ADU/s]}
& {\tiny [ADU/s]} 
& {\tiny [ADU/s]}    
&\\
\noalign{\vskip3pt}
\hline
\noalign{\vskip3pt}
V1 &{\it V} &2455542.05923 & 12.538& 13.705 & 0.001 & 32756.433& 5.278 & +194.010& 39.867&
0.9835\\
V1 &{\it V} &2455542.06303 & 12.531& 13.698 & 0.001 & 32756.433& 5.278 & +402.337& 36.651&
0.9932\\
{\vdots}   &  \vdots  & \vdots & \vdots & \vdots & \vdots   & \vdots & \vdots  & \vdots& \vdots \\
V2 &{\it V} & 2455542.05923& 14.830& 15.997 & 0.004 & 3897.998 & 5.372 & +92.541 & 13.824&
0.9835\\
V2 &{\it V} & 2455542.06303& 14.836& 16.004 & 0.004 & 3897.998 & 5.372 & +69.189 & 13.055
&0.9932\\
\vdots   &  \vdots  & \vdots & \vdots & \vdots & \vdots   & \vdots & \vdots  & \vdots& \vdots \\
V2 &{\it I}& 2455542.03472& 14.467& 15.543& 0.004& 5802.472 &  16.287 &+259.464& 19.695 &
0.9826\\
V2 &{\it I}& 2455542.03652& 14.462& 15.537& 0.004& 5802.472& 16.287&+289.454& 21.511&
0.9805\\
\vdots   &  \vdots  & \vdots & \vdots & \vdots & \vdots   & \vdots & \vdots  & \vdots& \vdots \\
\noalign{\vskip3pt}
\hline
\noalign{\vskip3pt}
\multicolumn{11}{p{12.5cm}}{Standard $M_{\rm std}$ and
instrumental $m_{\rm ins}$ magnitudes are listed in columns 4 and 5,
respectively, corresponding to the variable star in column~1. Filter and
epoch of mid-exposure are listed in columns 2 and 3, respectively. The
uncertainty on $m_{\rm ins}$ is listed in column 6, which also corresponds
to the uncertainty on $M_{\rm std}$. For completeness, we also list the
quantities $f_{\rm ref}$, $f_{\rm diff}$ and $p$ from Eq.(1) in columns 7,
9 and 11, along with the uncertainties $\sigma_{\rm ref}$ and $\sigma_{\rm
diff}$ in columns 8 and 10. This is an extract from the full table, which
is available from the {\it Acta Astronomica Archive}.}}

The standard minus the instrumental magnitudes indeed show a mild
dependence on the color, as can be seen in Fig.~1. We have transformed the
instrumental $\vv$ and $i$ into the standard {\it V} and {\it I} magnitudes
using the following equations:
\setcounter{equation}{2}
\begin{eqnarray}
V&=\vv+0.0782(\pm0.0092)(\vv-i)-1.2065(\pm0.0078),\\
I&=i+0.0271(\pm0.0111)(\vv-i)-1.0889(\pm0.0094).
\end{eqnarray}

Due to the lack of observations in the {\it I}-band for some bright
saturated stars, to calculate their standard {\it V} magnitudes, we have
adopted for them $(\vv-i)=0.5$~mag, which corresponds approximately to the
center of the RR~Lyr horizontal branch (HB).

Fig.~2 shows the RMS magnitude deviation in our {\it V} and {\it I}
light curves as a function of the mean magnitude. We achieve an RMS
scatter of $\approx10{-}20$~mmag in both the {\it V} and {\it I} filters
for stars brighter than 17~mag.
\begin{figure}[htb] 
\includegraphics[width=11cm]{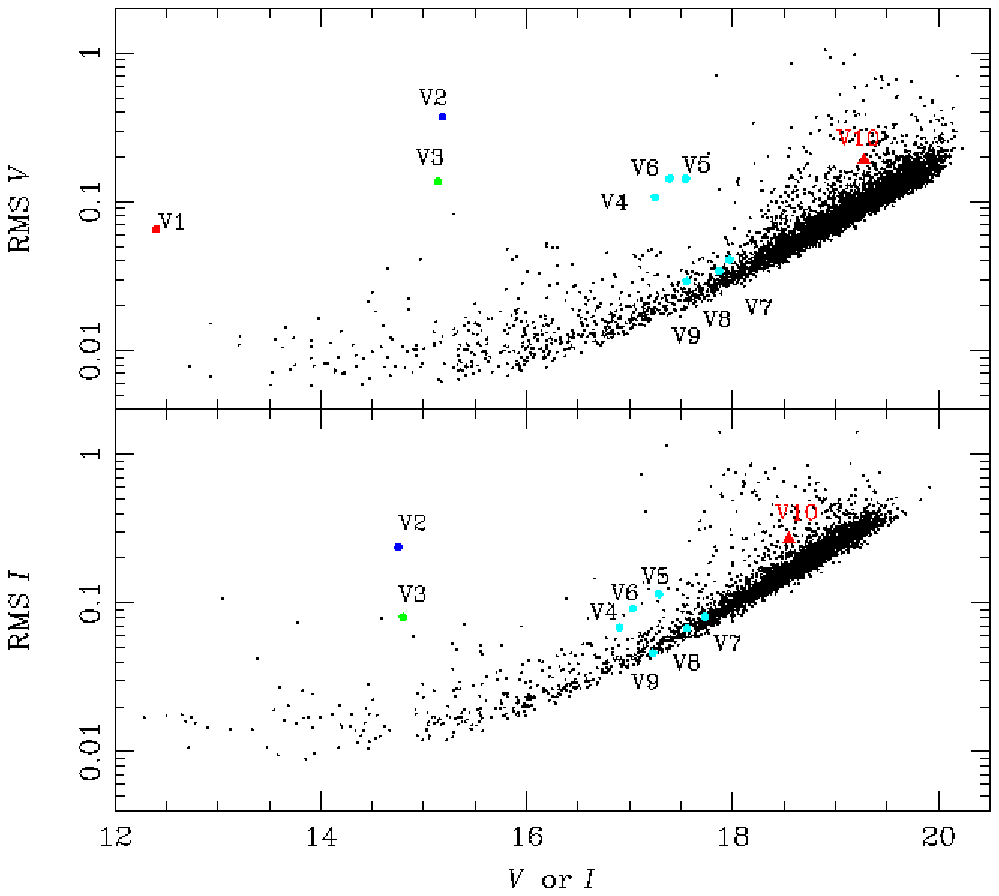}
\FigCap{The RMS magnitude deviations as a function of magnitude. The {\it 
upper and lower panels} correspond to the {\it V} and {\it I} light curves
respectively. The color coding is as follows: RRab star -- blue circle
(V2), RRc star -- green circle (V3), cyan circles are SX~Phe stars (V4--V9)
and red triangle is an eclipsing binary (V10). The long period variable V1
is not included in the {\it bottom panel} because it is saturated in the
{\it I} images in our collection.}
\end{figure}

All of our {\it V} and {\it I} photometry for the variable stars in the FoV
of our images of NGC~288 is reported in Table 2. Only a small portion of
Table~2 is given in the printed version of this paper, while the full table
is available in the electronic form from the {\it Acta Astronomica
Archive}.

\subsection{Astrometry}
A linear astrometric solution was derived for the {\it V} filter reference
image by matching $\approx250$ hand-picked stars with the UCAC3 star
catalogue (Zacharias \etal 2010) using a field overlay in the image display
tool GAIA (Draper 2000). We achieved a radial RMS scatter in the
residuals of $\approx0\zdot\arcs3$. The astrometric fit was then used to
calculate the J2000.0 celestial coordinates for all of the confirmed
variables in our field of view (see Table~3). The coordinates correspond to
the epoch of the {\it V} reference image, which pertains to the
heliocentric Julian day 2455542.07~d.
\vspace*{-7pt} 
\Section{Search for New Variable Stars}
\vspace*{-3pt} 
In this section we describe the use of our time-series {\it V} and {\it I}
photometry to search for new variables and to revisit the identifications,
periodicities, and light curves of the known variables.

\subsection{Search Methods}
\vspace*{-3pt} 
Ten variable stars listed in Table~3 all previously known, are identified
in the finding chart of Fig.~6. Their distribution is apparently peculiar
since they seem concentrated in a rather small off-centered region in the
FoV. Before we comment on the statistical significance of this distribution
we have conducted a search for new variables using our light curves. For
this purpose we have used several methods which have been successful in
previous studies in identifying and classifying variable stars in globular
clusters. Since these methods have already been described in detail in
earlier papers (\eg Arellano Ferro \etal 2013, Figuera Jaimes \etal 2013),
here we only summarize them briefly.
\renewcommand{\arraystretch}{0.96}
\MakeTableee{l@{\hspace{3pt}}l@{\hspace{1pt}}c@{\hspace{5pt}}c@{\hspace{5pt}}c@{\hspace{5pt}}c@{\hspace{5pt}}l@{\hspace{5pt}}l@{\hspace{1pt}}l@{\hspace{3pt}}l@{\hspace{5pt}}c@{\hspace{5pt}}c}{12.5cm}{General data for all of the confirmed 
variables in NGC~288 in the FoV of our images}
{\hline
\noalign{\vskip3pt}
Variable & Variable & $\langle V\rangle$ & $\langle I\rangle$ & $A_V$ & $A_I$ & \multicolumn{1}{c}{$P_1$} & HJD$_{\max}$ & \multicolumn{1}{c}{$P_2$} & RA & Dec.\\
Star ID  & Type     & {\tiny [mag]} & {\tiny [mag]} & {\tiny [mag]} &{\tiny
[mag]} & \multicolumn{1}{c}{\tiny [d]} & {\tiny (+2\,450\,000)} & \multicolumn{1}{c}{{\tiny [d]}} &{\tiny (J2000.0)} & {\tiny(J2000.0)}\\
\noalign{\vskip3pt}
\hline
\noalign{\vskip3pt}
V1  & SR     & 12.40~~& --     & 0.22~~& --   & --         & 5779.2366 & 103.$^a$ & 00\uph52\upm41\zdot\ups13 & $-26\arcd33\arcm27\zdot\arcs0$\\       
V2  & RRab   & 15.237 & 14.788 & 1.137 & 0.74 & 0.6777478  & 5842.2373 & 0.67775  & 00\uph52\upm46\zdot\ups69 & $-26\arcd34\arcm08\zdot\arcs1$\\
V3  & RRc    & 15.177 & 14.829 & 0.387 & 0.24 & 0.4301268  & 5871.1484 & 0.4302   & 00\uph52\upm40\zdot\ups27 & $-26\arcd32\arcm29\zdot\arcs3$\\               
V4  & SX Phe & 17.264 & 16.898 & 0.316 & 0.20 & 0.07907489 & 5842.3110 & 0.07907  & 00\uph52\upm42\zdot\ups85 & $-26\arcd34\arcm46\zdot\arcs1$\\    
V5  & SX Phe & 17.589 & 17.284 & 0.414 & 0.34 & 0.05106684 & 5868.2112 & 0.05107  & 00\uph52\upm45\zdot\ups03 & $-26\arcd33\arcm52\zdot\arcs7$\\ 
V6  & SX Phe & 17.400 & 17.028 & 0.462 & 0.30 & 0.06722082 & 5543.0555 & 0.06722  & 00\uph52\upm42\zdot\ups45 & $-26\arcd34\arcm55\zdot\arcs2$\\
V7  & SX Phe & 17.981 & 17.730 & 0.077 & --   & 0.03997368 & 5871.1232 & 0.03996  & 00\uph52\upm41\zdot\ups44 & $-26\arcd34\arcm00\zdot\arcs3$\\       
V8  & SX Phe & 17.875 & 17.555 & 0.061 & --   & 0.04652840 & 5871.2625 & 0.04653  & 00\uph52\upm44\zdot\ups32 & $-26\arcd34\arcm00\zdot\arcs3$\\
V9  & SX Phe & 17.558 & 17.223 & 0.04  & --   & 0.03936761 & 5842.2475 & 0.03937  & 00\uph52\upm42\zdot\ups94 & $-26\arcd34\arcm10\zdot\arcs3$\\      
V10 & E      & 19.275 & 18.545 & $\approx0.5$ & $\approx0.7$ & 0.4387538 & 5542.0865$^b$ & 0.43875 & 00\uph52\upm47\zdot\ups91 & $-26\arcd33\arcm02\zdot\arcs5$ \\
\noalign{\vskip3pt}
\hline
\noalign{\vskip5pt}
\multicolumn{11}{p{12.5cm}}{The best previous period estimates for each 
variable from Kaluzny \etal (1997) are reported in column~9 ($P_2$) for
comparison with our refined periods in column~7 ($P_1$). The period
uncertainties are within the last significant digit.\newline  $^a$ from Oosterhoff
(1943), $^b$ Time of minimum light.}}

\begin{figure}[htb]
\vglue-1mm
\centerline{\includegraphics[width=9.3cm]{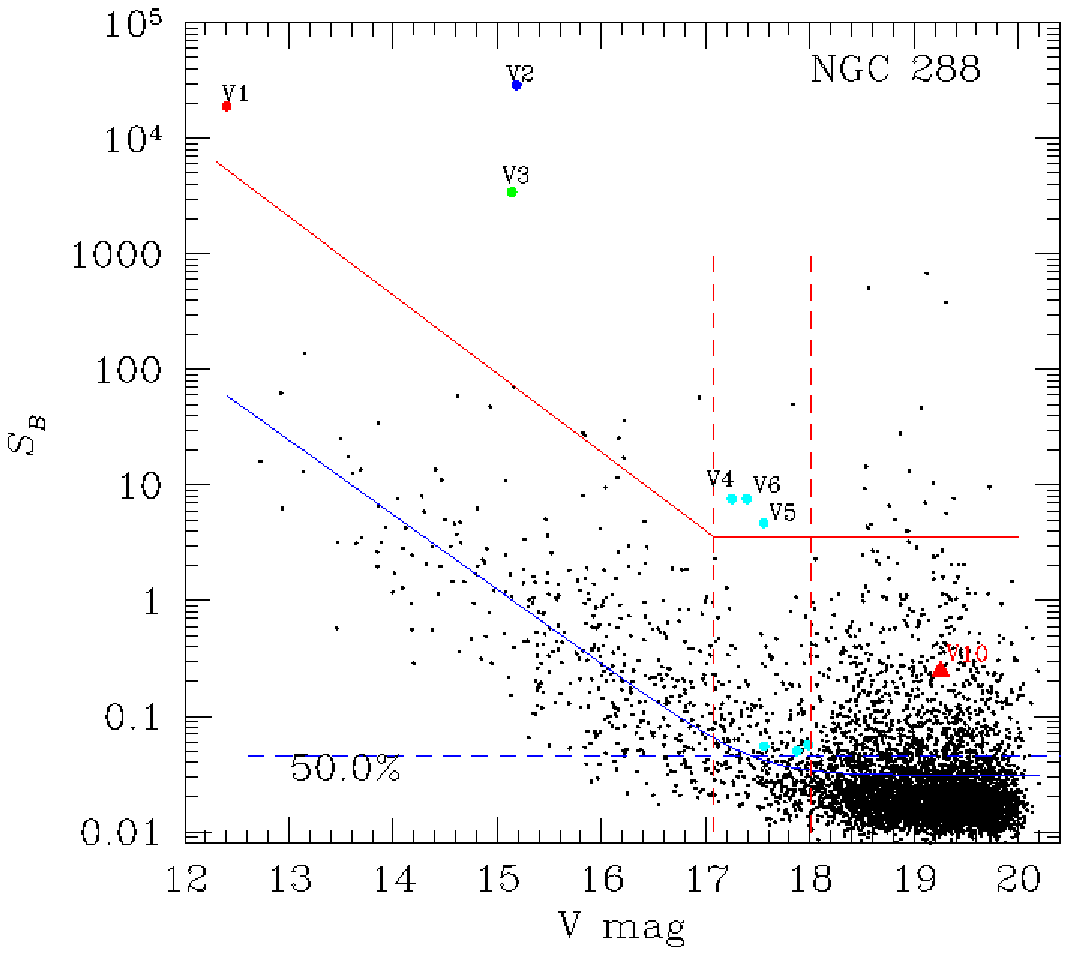}}
\vskip-5mm
\FigCap{Distribution of the $\cal S_B$ statistic as a function of mean {\it V}
magnitude for 5525 stars measured in the {\it V} images of NGC~288. The
colored symbols for variable stars are as described in the caption of
Fig.~2. The blue lines represent the median (50\%) percentile for the
simulated light curves and the real distribution of $\cal S_B$. The red
solid line is the threshold set by eye above which one expects to find true
variable stars. See the discussion in Section~3.1. The two vertical dashed
red lines correspond to the magnitude limits set for the Blue Straggler
region in the CMD.}
\end{figure}
\begin{figure}[p] 
\centerline{\includegraphics[width=9.3cm]{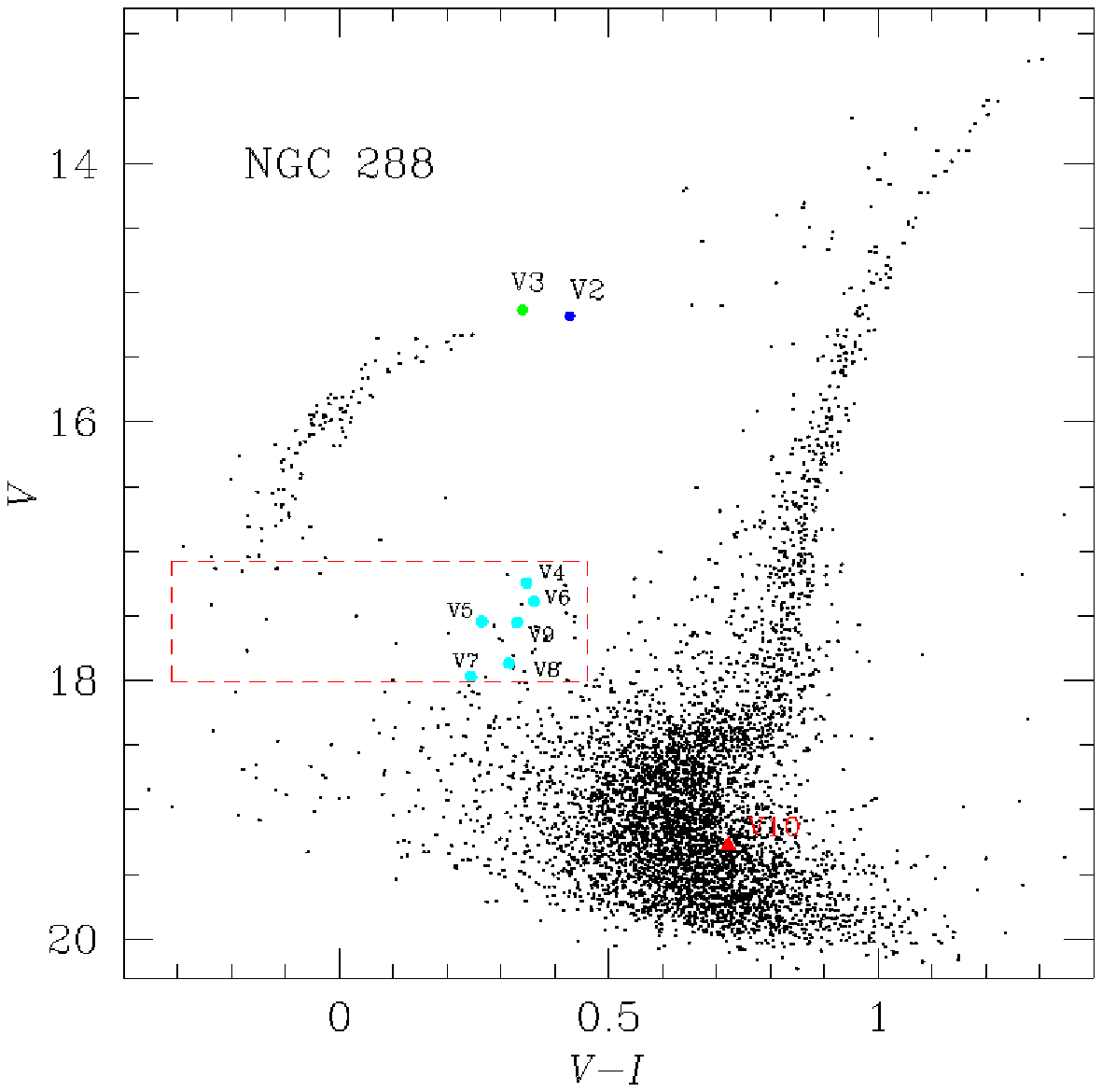}}
\FigCap{Color--magnitude diagram of NGC~288. The colored symbols are as
in the caption of Fig.~2. The long period variable V1 is not included
because it is saturated in the {\it I} images in our collection.}
\vskip9mm
\centerline{\includegraphics[width=9.5cm]{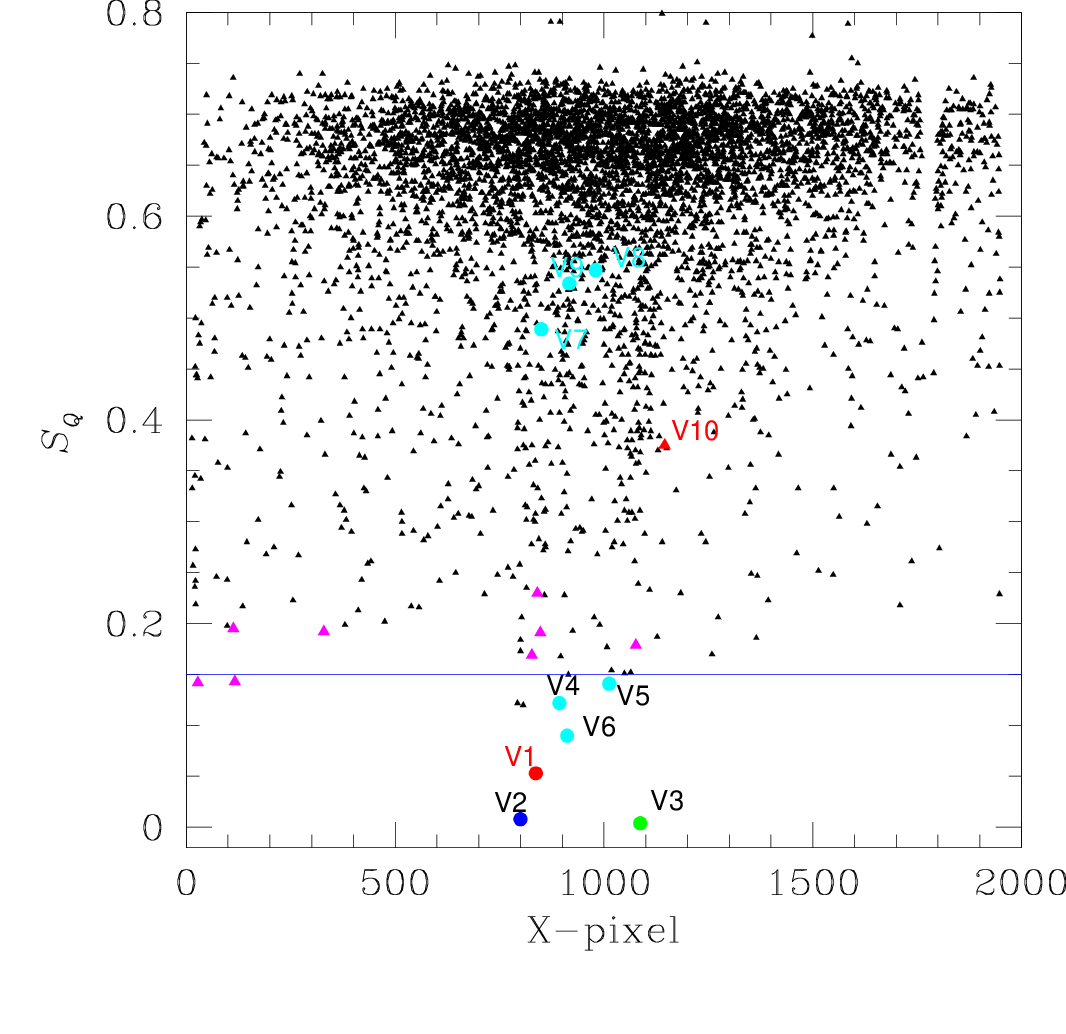}}
\vspace*{-5mm}
\FigCap{Minimum value of the string-length parameter $S_Q$ calculated for 
the 5525 stars with {\it V} light curves \vs the CCD $x$-coordinate. The
colored symbols are as described in the caption of Fig.~2.}
\end{figure}

Firstly, we have defined a variability statistic $\cal S_B$ as:
$${\cal S_B}=\frac{1}{NM}\sum\limits_{i=1}^{M}\left(\frac{r_{i,1}}{\sigma_{i,1}}+\frac{r_{i,2} }{
\sigma_{i,2}}+\dots+\frac{r_{i,k_i}}{\sigma_{i,k_i}}\right)^2\eqno(5)$$
where $N$ is the total number of data points in the light curve and $M$ is
the number of groups of time-consecutive residuals of the same sign from
the inverse-variance weighted mean magnitude. The residuals $r_{i,1}$ to
$r_{i,k_i}$ form the $i$th group of $k_i$ time-consecutive residuals of the
same sign with corresponding uncertainties $\sigma_{i,1}$ to
$ \sigma_{i,k_i}$. Fig.~3 shows the distribution of the $\cal S_B$
statistic as a function of mean magnitude for the 5525 light curves for the
stars in the {\it V} images.

As in the paper by Figuera Jaimes \etal (2013) we calculated $10^6$
randomly generated light curves (their Eq.~4) and computed their $\cal S_B$
values.  The median of the distribution, or 50\% percentile is indicated by
the dashed blue line in Fig.~3. The fit to the real $\cal S_B$ values is
shown as a solid blue line for $V>18$~mag and we notice that the real and
the simulated values are very close, \ie the simulations are a good
description of the statistical noise for these faint stars. For brighter
stars however ($V<18$~mag) systematic errors dominate and $\cal S_B$
increases logarithmically with stellar magnitude.  As before, the solid
blue line is the fit to the real distribution of $\cal S_B$.  Then, guided
by the distribution of the known variables, we defined by eye a variability
detection threshold indicated by the solid red line in Fig.~3. RR~Lyr stars
are generally easily identified by this method as they have substantially
larger values of $\cal S_B$ among stars of their magnitude range. The same
is true of the SR long-term variables. Short-period small-amplitude SX~Phe
stars might however escape detection by this approach. As it can be seen in
Fig.~3 the three large amplitude SX~Phe stars (V4, V5 and V6) are found
clearly above the threshold. However the three low amplitude ones (V7, V8,
and V9) and the known binary star (V10) are buried below in the cloud of
otherwise non-variable stars.

Only seven stars brighter than 17 mag and one in the BS region (between the
two dashed red lines) have $\cal S_B$ larger than the detection
threshold. We have explored their {\it V} and {\it I} light curves in
detail. No convincing signs of true variability were found. For several of
them their large $\cal S_B$ is explained by their proximity to an authentic
known variable star or to a poorly subtracted bright star, hence their
difference fluxes suffer correlated systematic errors.

A second strategy that we applied was the string-length method (Burke,
Rolland and Boy 1970, Dworetsky 1983) to each light curve to determine the
period and a normalized string-length statistic $S_Q$. In Fig.~5 we plot
the minimum $S_Q$ value for each light curve as a function of their
corresponding CCD $x$-coordinate. The known variables are plotted with the
colored symbols as described in the caption. The horizontal blue line is
not a statistically defined threshold but again, set by eye, as an upper
limit to the majority of the known variables. In fact this method could
recover variables V1 to V6 but also fails in detecting the low-amplitude
short-period variables V7 to V10. The eight stars spotted before as having
large $\cal S_B$ values are plotted with purple triangles and we note that
only two would pass the $S_Q$ threshold requirement. There are six other
stars below the $S_Q$ threshold line. However, as before, the exploration
of their light curves did not reveal any true variability.

Finally, we have followed a third approach to identify variables in the
field of our images by detecting PSF-like peaks in a stacked image built
from the sum of the absolute valued difference images normalized by the
standard deviation in each pixel as described by Bramich \etal (2011). This
method allowed us to confirm the variability of all ten known variables in
Table~3 but no new variables emerged.

In conclusion, we did not find any new variable stars in our light curve
collection using the above three methods. We believe that our search for
variable stars with continuous variations (\ie not eclipsing binaries) is
fairly complete down to $V \approx18$~mag for amplitudes larger than
$\approx0.05$~mag and periods between $\approx0.02$~d and a few hundred
days.

\begin{figure}[p]
\vglue-4mm
\centerline{\includegraphics[width=13cm]{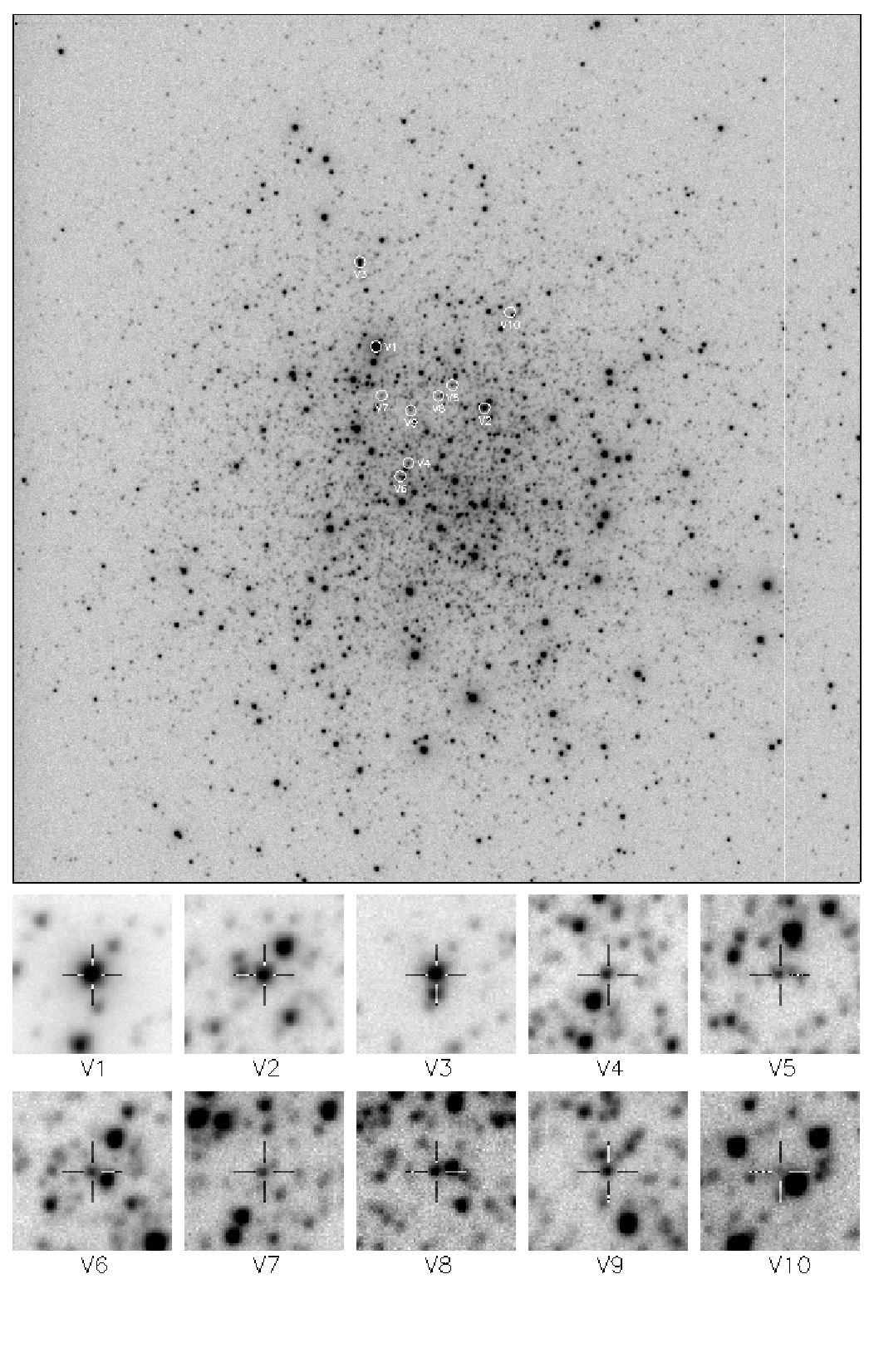}}
%\centerline{\includegraphics[width=10cm, bb=50 405 555 1150]{fig6.ps}}
\FigCap{Finding charts constructed from our Hanle {\it V} reference image. 
North is up and East is to the right. The cluster image is $9.62\times9.87$~arcmin$^{2}$, 
and the image stamps are of size $23.7\times23.7$~arcsec$^{2}$. Each variable lies at 
the center of its corresponding image stamp and is marked by a cross-hair.}
\end{figure}
\vspace*{9pt}
\subsection{Period Determination and Refinement}
\vspace*{5pt}
We have combined our {\it V} light curves with those of Kaluzny \etal
(1997), taken between 1990 and 1992, and of Kaluzny (1996), taken in 1995,
to re-calculate the periods. Since more than twenty years have passed
between the two data sets, their combination leads to substantially refined
periods.

We have noticed small zero point differences of 0.02~mag to 0.08~mag
between our light curves and those of Kaluzny \etal (1997). This is to be
expected since the error in the reference flux affects all photometric
measurements for a single star in the same way. The procedure to estimate
the refined period was as follows: first we estimated the period using the
combined data sets and the program {\sf period04} (Lenz and Breger 2005)
and used this to phase the light curves. The necessary magnitude shift was
applied to the data from Kaluzny (1996) and Kaluzny \etal (1997) and then
the leveled combined light curve was period analyzed using a dense scan of
the string-length method within a short period range around the initial
estimate of the period. The new periods are given in column 7 of Table~3
and have been used to phase the light curves shown in Figs.~7
and~9. Variations of the period within the last significant digit can spoil
the phasing of the light curve, which sets the uncertainty of the
periods. As a reference we list the periods from Kaluzny \etal (1997) in
column~9.

\Section{RR Lyrae Stars}
\subsection{Physical Parameters from Light Curve Fourier Decomposition}
An estimation of [Fe/H], $M_V$, and $T_{\rm eff}$ for a given RR~Lyr star
can be obtained by Fourier decomposing the light curve into its harmonics
as
$$m(t)=A_0+\sum_{k=1}^NA_k\cos\left(\frac{2\pi}{P}k(t-E)+\phi_k\right)\eqno(6)$$
where $m(t)$ are magnitudes at time $t$, $P$ the period and $E$ the
epoch. A linear minimization routine is used to fit the data with the
Fourier series model, deriving the best fit values of the amplitudes $A_k$
and phases $\phi_k$ of the sinusoidal components. From the amplitudes and
phases of the harmonics in Eq.(6) the Fourier parameters
$\phi_{ij}=j\phi_i-i\phi_j$ and $R_{ij}=A_i/A_j$ are computed.

These Fourier parameters can be used in the semi-empirical calibrations of
Jurcsik and Kov\'acs (1996), for RRab stars, and Morgan, Wahl and
Wieckhorts (2007), for RRc stars, to obtain ${\rm [Fe/H]}_{ZW}$ on the Zinn
and West (1984) scale. The absolute magnitude M$_V$ can be estimated from
the calibrations of Kov\'acs and Walker (2001) for RRab stars and of
Kov\'acs (1998) for the RRc stars. The effective temperature $T_{\rm eff}$
was estimated using the calibration of Jurcsik (1998). For brevity we do
not present here the specific equations but they can be found in a recent
paper (Arellano Ferro
\etal 2013).

The mean magnitudes $A_0$, and the Fourier light curve fitting parameters
for V2 (RRab) and V3 (RRc) in the {\it V} filter are listed in Table~4. The
absolute magnitude $M_V$ was converted into $\log L/\LS=-0.4(M_V-M_{\rm
bol}+BC$). The bolometric correction was calculated using the formula $BC=
0.06{\rm [Fe/H]}_{ZW}+0.06$ given by Sandage and Cacciari (1990). We
adopted the value $M^\odot_{\rm bol}=4.75$~mag.
\begin{figure}[htb] 
\includegraphics[width=12.5cm]{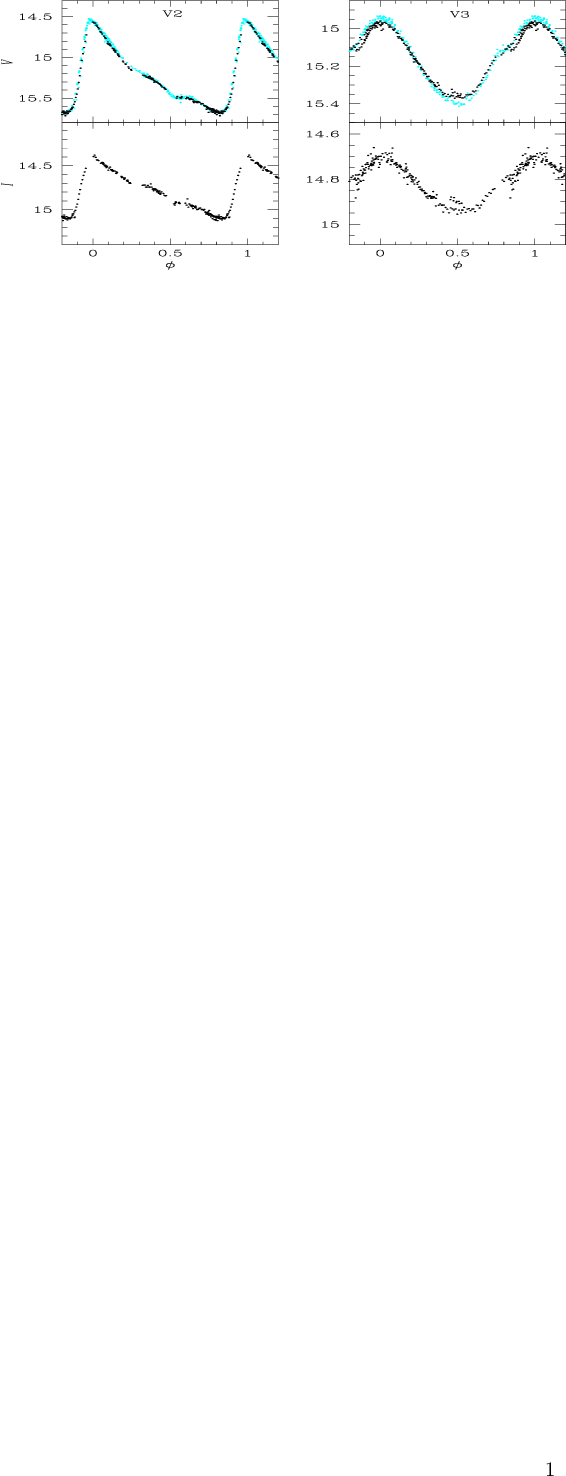}
\FigCap{Standard magnitude {\it V} and {\it I} light curves of the RR~Lyr
stars V2 and V3 phased with the periods listed in Table~3. Black points
represent Hanle data from the present work. Blue points represent {\it V}
data from Kaluzny \etal (1997). Typical uncertainties in {\it V} and {\it
I} are $\approx6$~mmag.}
\end{figure}

\MakeTable{l@{\hspace{5pt}}c@{\hspace{7pt}}c@{\hspace{7pt}}c@{\hspace{7pt}}c@{\hspace{7pt}}c@{\hspace{7pt}}c@{\hspace{7pt}}c@{\hspace{7pt}}c@{\hspace{7pt}}c@{\hspace{7pt}}c}{12.5cm}{Fourier coefficients $A_k$ for $k=0,1,2,3,4$, and phases $\phi_{21}$,
$\phi_{31}$ and $\phi_{41}$, for V2 (RRab) and V3 (RRc) in NGC~288} 
{\hline
\noalign{\vskip3pt}
Variable     & $A_0$    & $A_1$   & $A_2$   & $A_3$   & $A_4$   &
$\phi_{21}$ & $\phi_{31}$ & $\phi_{41}$ 
& $N$   & $D_{\rm m}$ \\
  ID     & [{\it V} mag]  & [{\it V} mag]  &  [{\it V} mag] & [{\it V} mag]
  & [{\it V} mag] & & & & & \\
\noalign{\vskip3pt}
\hline
\noalign{\vskip3pt}
V2 & 15.237(1)& 0.401(2) &0.204(2) &0.138(2)& 0.065(2)&4.257(12)& 8.513(18)&
6.805(34)& 9  & 3.5\\
V3 & 15.177(1)& 0.188(2) &0.008(1) & 0.014(1) & 0.011(1) &2.023(170) &5.115(99)
&3.396(130) & 6  & --\\
\noalign{\vskip3pt}
\hline
\noalign{\vskip5pt}
\multicolumn{11}{p{12.5cm}}{The numbers in parentheses indicate
the uncertainty on the last decimal place. Also listed are the number of
harmonics $N$ used to fit the light curve of each variable and the deviation 
parameter $D_{\rm m}$ (see Section~4.1).}       
}

The resulting physical parameters are given in Table~5. The average
metallicity obtained from the RRab and RRc stars is ${\rm
[Fe/H]}_{ZW}=-1.62\pm0.02$ which can be converted to the new scale defined
by Carretta \etal (2009) using UVES spectra of RGB stars in globular
clusters by
$${\rm [Fe/H]}_{UVES}=-0.413 +0.130{\rm [Fe/H]}_{ZW}-0.356{\rm [Fe/H]}_{ZW}^2.\eqno(7)$$
We find ${\rm [Fe/H]}_{UVES}=-1.56\pm0.03$. No previous estimates of [Fe/H]
from Fourier decomposition of the RR~Lyr light curves exist for this
cluster.
\MakeTable{lccccccc}{12.5cm}{Physical parameters for V2 (RRab) and V3 (RRc) stars}
{\hline 
\noalign{\vskip3pt}
Star& ${\rm [Fe/H]}_{ZW}$ & $M_V$ & $\log(L/\LS)$ & $\log T_{\rm eff}$  &
$M/\MS$ & $R/\RS$\\
\noalign{\vskip3pt}
\hline
\noalign{\vskip3pt}
V2 &$-1.642(17)$&0.376(3)&1.750(1)&3.800(8)& 0.74(7)&6.04(1)\\
V3 &$-1.59(24)$ &0.579(7)&1.668(1)$^a$&3.856(1)& 0.38(1)$^a$&5.57(9)\\
\noalign{\vskip3pt}
\hline
\noalign{\vskip5pt}
\multicolumn{8}{p{9cm}}{$^a$These values depend on $M_V$ whose
peculiarity for V3 is discussed in Section~4.2. The numbers in parentheses
indicate the uncertainty on the last decimal place and have been calculated
as described in the text.}}

Zinn (1980) made an early estimation of iron content of NGC~288 from
integrated photometry in the $Q_{39}$ index. He found ${\rm [Fe/H]}=-1.61$.
Zinn and West (1984) summarized previous estimates of [Fe/H] given in
earlier papers in a variety of metallicity scales (their Table~5). They
report the weighted average ${\rm [Fe/H]}_{ZW}=-1.40$ in their new
scale. The high resolution spectroscopic value of [Fe/H], derived by
Carretta and Bragaglia (1998) from two giant cluster member stars is $-1.07$
in their own scale which is equivalent to ${\rm [Fe/H]}_{ZW}=-1.40$ in the
Zinn and West (1984) scale. Our calculation from two independent
calibrations for RRab and RRc stars of the Fourier decomposition parameters
average ${\rm [Fe/H]}_{ZW}=-1.62\pm0.02$ (statistical) $\pm0.14$
(systematic), favors a lower iron content.

\subsection{Distance to NGC~288 from the RR~Lyr Stars}
The $M_V$ value calculated for the RRab and RRc stars in Table~5 can be
used to estimate the true distance modulus. Given the fact that these $M_V$
values come from independent calibrations for the RRab and RRc stars, with
their own systematic uncertainties, we should consider the distances as
derived from them as two independent estimations. We adopted
$E(B-V)=0.03$~mag (Buonanno \etal 1984, Harris 1996). We find the true
distance moduli of $14.768\pm0.003$~mag and $14.505\pm0.008$~mag using the
RRab and RRc stars respectively. The uncertainties are only the internal
errors which are small. These moduli correspond to the distances 8.99~kpc
and 7.96~kpc with corresponding internal and systematic errors of $\approx
0.01$~kpc and $\approx0.17$~kpc respectively. The distance to NGC~288
listed in the catalogue of Harris (1996, 2010 edition) is 8.9~kpc.

\begin{figure}[htb] 
\centerline{\includegraphics[width=9cm]{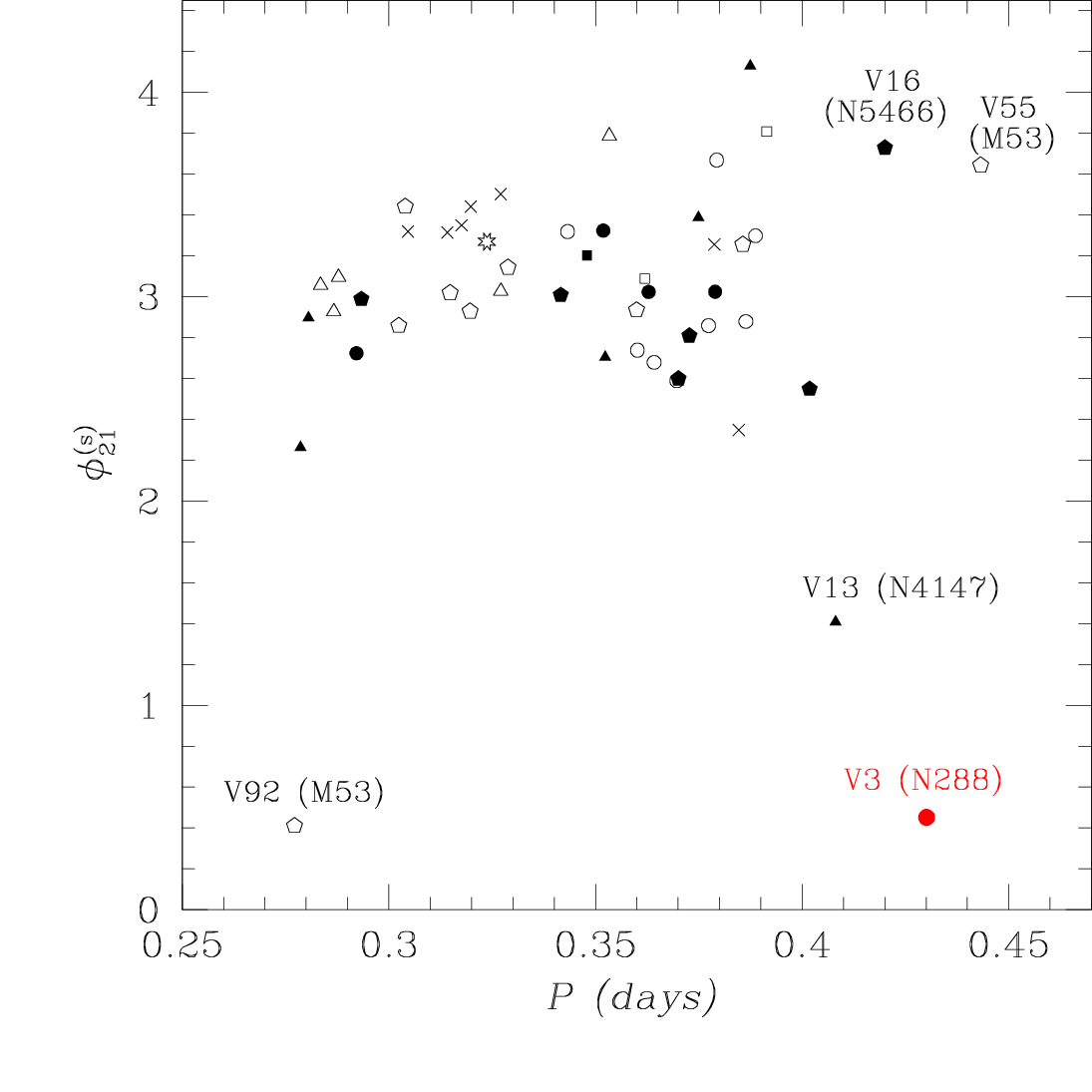}}
\FigCap{Distribution of RRc stars from a family of clusters in the $\phi^{(s)}_{21}
{-}P$ plane. Symbol codes and data sources are: filled triangles NGC~4147
(Arellano Ferro \etal 2004), open circles M15 (Arellano Ferro, Garc\'ia
Lugo and Rosenzweig 2006), M2 open squares (L\'azaro \etal 2006), NGC~5466
filled pentagons (Arellano Ferro \etal 2008), NGC 5053 filled circles
(Arellano Ferro, Giridhar and Bramich 2010), M72 open triangles (Bramich
\etal 2011), M53 open pentagons (Arellano Ferro \etal 2011), M79 starred
circle (Kains \etal 2012), M30 filled square (Kains \etal 2013), M9 crosses
(Arellano Ferro \etal 2013). In the long period range V3 in NGC 288 and V13
in NGC 4147 stand out from the distribution. The short period star V92 in
M53 is also peculiar. See text in Section~4.2 for a discussion.}
\end{figure}

The distance obtained from the RRc star, V3, is discrepant when compared
with that from the RRab star and the generally accepted distance for the
cluster. In Fig.~8 we plot $\phi^{(s)}_{21}$ \vs\ $P$, which are the key
parameters for the calculation of $M_V$ in RRc stars (see Eq.~13 of
Arellano Ferro \etal 2013), for a group of RRc stars in several globular
clusters listed in the caption of Fig.~8. The corresponding data have been
taken from recent publications of our working group on those clusters. V3
is a long period RRc star and its $\phi^{(s)}_{21}$ value is very small,
these two facts highlight the star as peculiar. The other stars that stand
out from the distribution of RRc stars are the long period V13 in NGC~4147
and the short period variable V92 in M53. V13 was found by Arellano Ferro
\etal (2004), to exhibit amplitude variations, probably due to the Blazhko
effect, while V92 shows an unusual low amplitude (Arellano Ferro \etal
2011). We have noticed in Fig.~7 the amplitude difference in the light
curves of V3 from Kaluzny \etal (1997) data and that from the present
work. While we remark that this amplitude difference may be an artefact
from the reduction processes of both data sets, or that the presence of the
Blazhko effect cannot be ruled out, with the present data we cannot
identify the reason for the peculiar position of V3 in the $\phi^{(s)}_{21}
{-}P$ plane. Due to these peculiarities, we do not give any weight to the
distance suggested by~V3.

\subsection{RR Lyr Masses and Radii}
Given the period, luminosity and temperature for each RR~Lyr star, its
mass and radius can be estimated from the equations: $\log M/\MS=
16.907-1.47\log P_F+1.24\log(L/\LS)-5.12\log T_{\rm eff}$ (van Albada and
Baker 1971) and $L=4\pi R^2\sigma T^4$ respectively.  The masses and radii
are given in Table~5 in solar units. Given the peculiarity of $M_V$ for V3
(see Section~4.2), $\log(L/\LS)$ and $\log M/\MS$ should be considered with
caution for this star.

\Section{SX Phoenicis Stars}
Six SX~Phe stars are known in NGC~288, three of large amplitude (V4, V5 and
V6) and three of low amplitude (V7, V8, V9). We have searched the light
curves of all stars in an arbitrarily defined blue straggler region in the
CMD of Fig.~4 delimited by the dashed red lines. As discussed in
Section~3.1 all our approaches to finding variable stars failed to reveal
any convincing new variables.

\begin{figure}[htb]
\includegraphics[width=12.5cm]{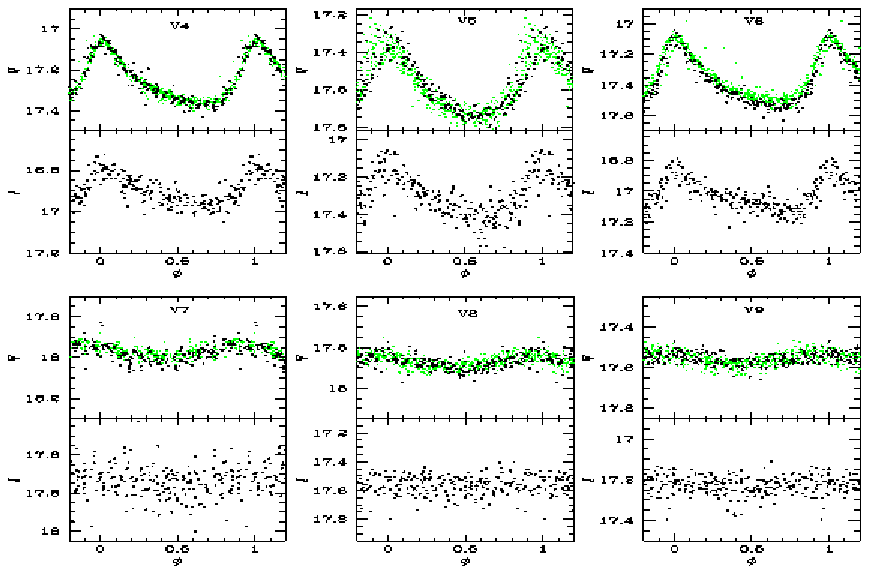}
\FigCap{Light curves of the SX~Phe stars. Green symbols are from Kaluzny
(1996) and Kaluzny \etal (1997) and black symbols from the present
work. Typical uncertainties are 0.03~mag and 0.05~mag for {\it V} and {\it
I} respectively.}
\end{figure}
The light curves of the six known SX~Phe are shown in Fig.~9 phased with
the refined periods and epochs listed in Table~3. We have included the data
from Kaluzny (1996) and Kaluzny et al. (1997) (green symbols). Our {\it I}
light curves are also shown in the bottom panels.

\begin{figure}[htb] 
\centerline{\includegraphics[width=8.5cm]{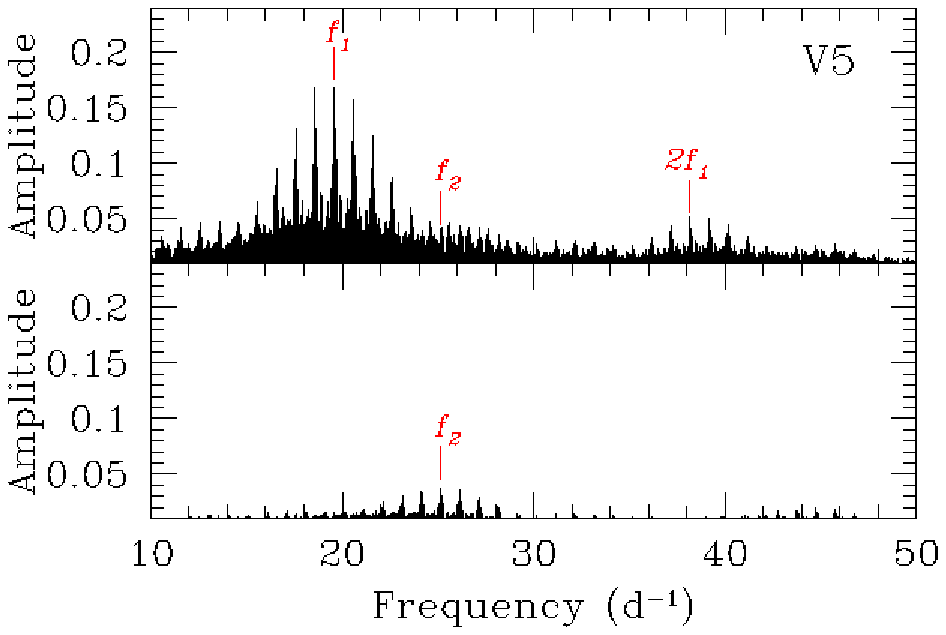}}
\vskip3pt
\FigCap{Frequency spectra of V5 calculated from the {\it V} light curve of Fig.~9.
The primary frequency $f_1$, its alias at 2$f_1$ and secondary frequency
$f_2$ are labelled. Since $f_1/f_2=0.779$ we interpret this as being a
double mode pulsator with the fundamental and the first overtones being
excited.}
\end{figure}
It is worth noting that V5 has a larger dispersion than V4 and V6 despite
being of similar magnitudes. We searched for a secondary frequency by
prewhitening the primary frequency $f_1=19.5821790\pm0.0000016$~d$^{-1}$
and its aliases. In Fig.~10 we note the presence of a secondary frequency
at $f_2=25.1482392\pm0.0000054$~d$^{-1}$. The ratio $f_1/f_2= 0.779$ lead
us to identify $f_1$ and $f_2$ with the fundamental and the first overtone
radial modes respectively.

\begin{figure}[htb] 
\centerline{\includegraphics[width=7.7cm]{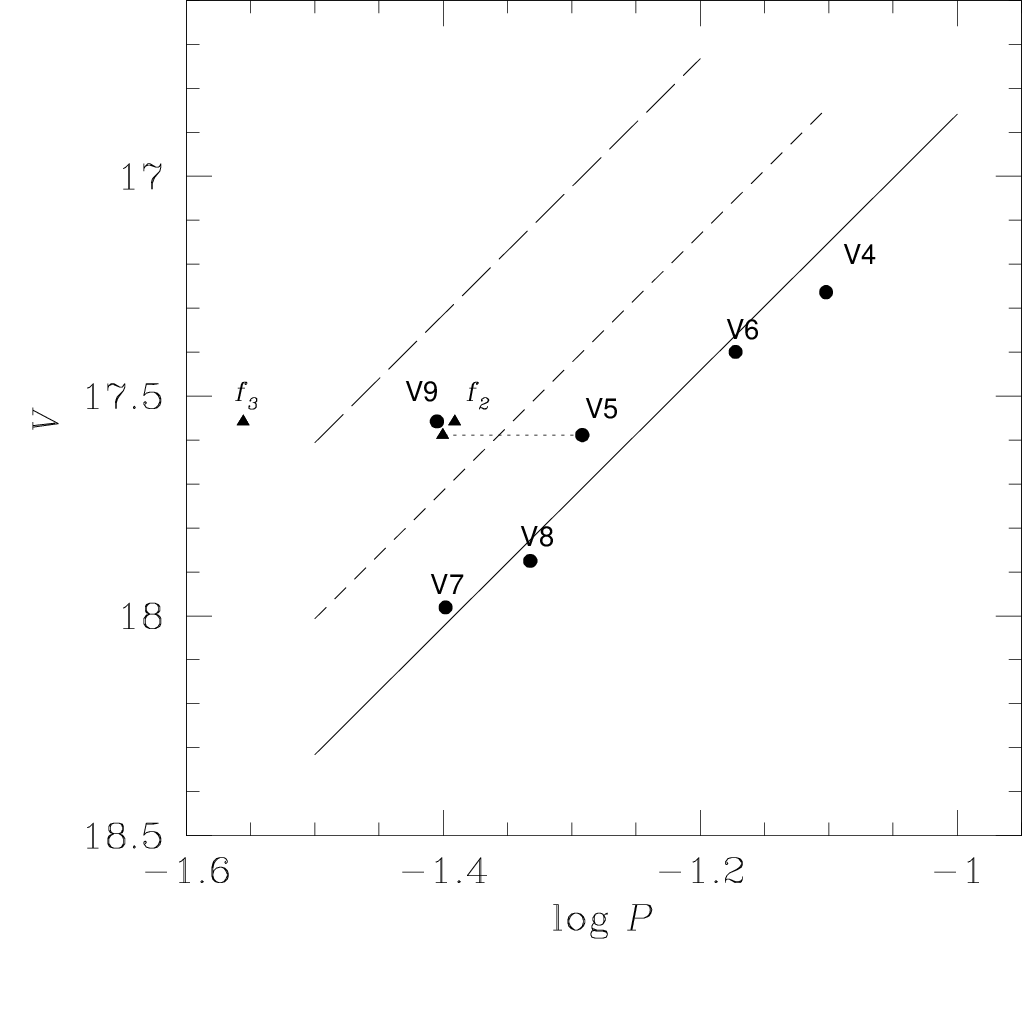}}
\FigCap{P--L relation for SX~Phe stars. The solid line corresponds to the SX~Phe P--L
relation of Arellano Ferro \etal (2011) derived in M53 scaled to a distance
of 8.9~kpc. Short and long dashed lines represent the loci of the
corresponding first and second overtone respectively. For V5 we represent
the fundamental mode (filled circle) and the first overtone (black
triangle) modes joined by the dotted line. For V9 we indicate the positions
of secondary frequencies $f_2$ and $f_3$ which are interpreted as a
non-radial mode and the second overtone, respectively.}
\end{figure}
Fig.~11 shows P--L diagram for the SX~Phe stars. Except for V9, the known
SX~Phe define a linear progression similar to the SX~Phe in other globular
clusters (\eg M53 Arellano Ferro \etal 2011 and Jeon \etal 2003, and M55
Pych \etal 2001). We adopted the SX~Phe P--L relation derived by Arellano
Ferro \etal (2011) in M53, $M_V=-2.916\log P-0.898$, to calculate $M_V$ and
hence the distance for each SX~Phe star. The average distance (excluding
V9) is $8.9\pm0.3$~kpc. The solid line in Fig.~11 is the $V{-}\log P$
relation corresponding to a distance of 8.9~kpc. As it is seen the match
with the SX~Phe distribution is very good. V5 lies slightly above the
relation but we have noted that it is blended with a fainter star. There is
a clear indication that V9 is pulsating in an overtone mode.

The above independent estimation of the distance is in excellent agreement
with the distance derived from the Fourier decomposition of the RRab star
V2 and with the generally adopted distance (8.9~kpc Harris 1996).

\begin{figure}[htb] 
\centerline{\includegraphics[width=9cm]{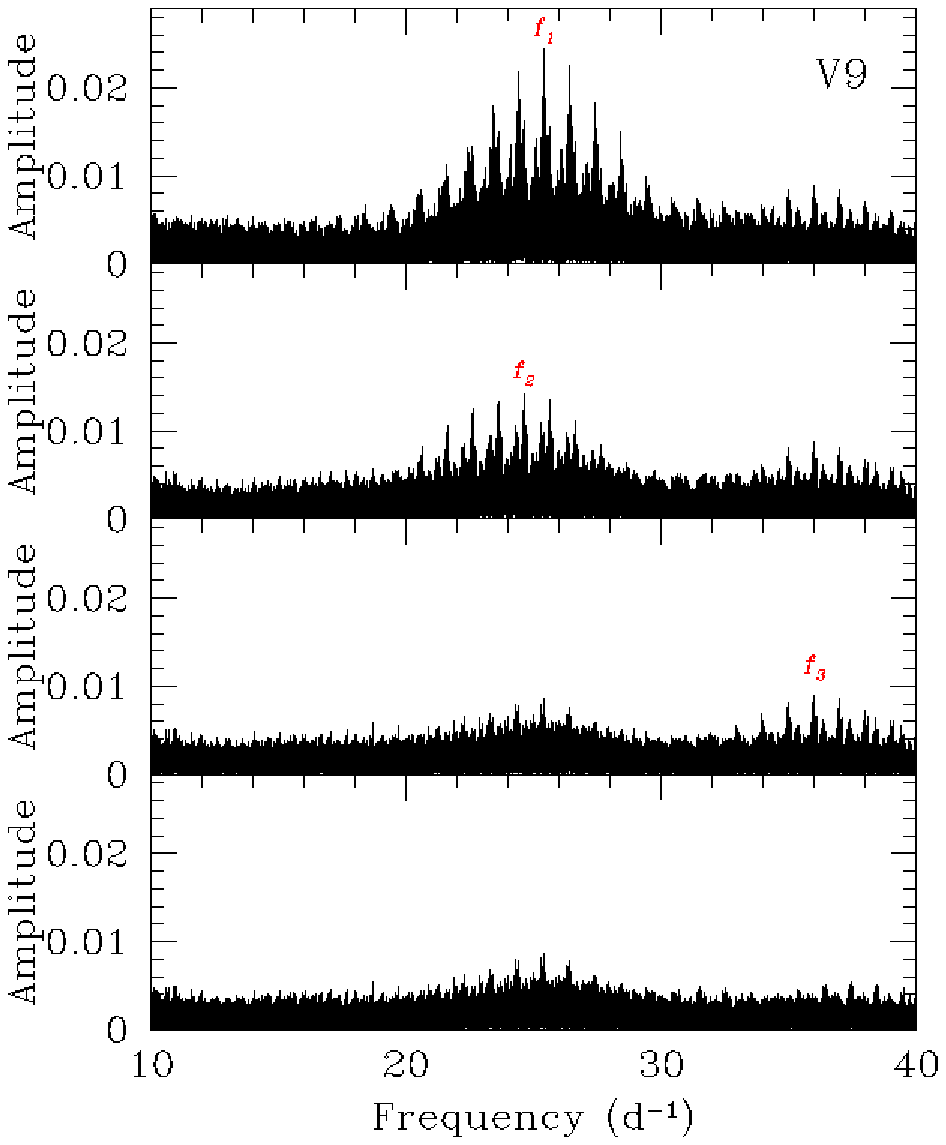}}
\vskip3pt
\FigCap{Frequency spectra of V9 calculated from the {\it V} light curve of 
Fig.~9. The secondary frequencies $f_2$ and $f_3$ are labelled and discussed 
in Section~5.}
\end{figure}
Then the question arises: if all known SX~Phe stars in NGC~288 pulsate in
the fundamental mode (except V9), why do some have a very large amplitude
(V4, V5, V6) and some a very small one (V7 and V8)? We explore the
possibility of more than one mode being excited in the small amplitude
stars. Using the frequency finding program {\sf period04} (Lenz and Breger
2005) we have prewhitened the main frequency, and in all cases except V9 we
found no signs of secondary frequencies. Therefore, V9 deserves special
attention. In Fig.~12 the frequency spectra calculated on the {\it V} light
curves are shown. The top panel corresponds to the original data, where the
main frequency $f_1=25.4015993$~d$^{-1}$ agrees well with the period found
from the string-length method (Table~3). The middle panel corresponds to
the spectrum after prewhitening the main frequency. The residuals show a
frequency of $f_2=24.61423$~d$^{-1}$. As the residual signal is substantial
a second prewhitening was performed to find a frequency of
$f_3=35.95981$~d$^{-1}$. The ratio $f_1$/$f_3$=0.706 is a bit off the
canonical value 0.783 for the fundamental and first overtone modes
(Santolamazza \etal 2001, Poretti \etal 2005). After removing $f_3$ the
signal virtually disappears; however some signal remains around $f_1$,
which probably indicates that we have not fully succeeded in determining
and removing the primary frequency, which given the scatter in the light
curve is probably not surprising.  On the other hand our frequency ratio of
0.706 is very close to the ratio between first and second overtone which is
0.729. In Fig.~11 we have also included the positions of V9 corresponding
to the frequencies $f_2$ and $f_3$ (black triangles). If one identifies
$f_1$ and $f_3$ with the first and second overtone frequencies
respectively, this strongly suggests that V9 is a double-mode SX~Phe
pulsating simultaneously in the first and second overtones while $f_2$
probably corresponds to a non-radial mode.

However, it is clear that this is a multifrequency variable, which accounts
for its very small amplitude. The fact that we did not find secondary
frequencies in the spectra of the other two small amplitude SX~Phe stars,
V7 and V8, is intriguing.

\Section{On the Variability of V1 and V10}
These two stars are a bright long period variable and a faint contact
binary respectively. Here we offer some comments on their variability and
periods.

\begin{figure}[htb]
\centerline{\includegraphics[width=9cm]{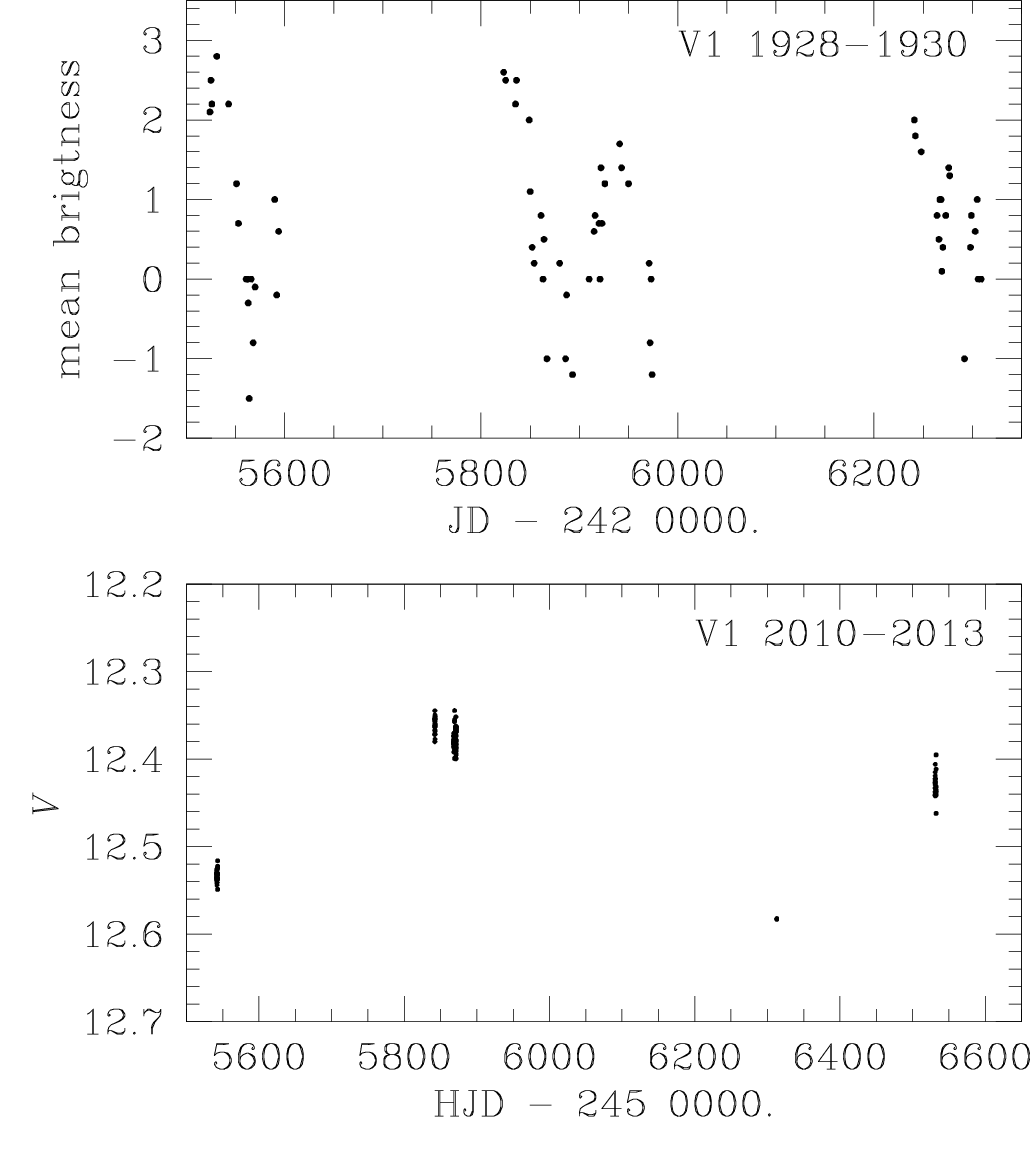}}
\FigCap{Light curves of the SR star V1. {\it Top panel:} the 
differential mean brightness from Table~1 of the discovery paper
(Oosterhoff 1943). A period analysis of these data confirms the 103d period
reported by Oosterhoff. {\it Bottom panel:} the variations from our
{\it V} photometry.}
\end{figure}
{\it V1}. The variability of this star was discovered by Oosterhoff (1943)
on photographic plates taken between 1928 and 1930. Despite the time
elapsed, the phase coverage in Oosterhoff's light curve continues to be the
best available. In Fig.~13 we plot the light curves from 1928--1930 and
2010--2013. A period analysis exclusively using the old data gives a period
of 103~d, as reported by Oosterhoff. Unfortunately it is not very clear how
Oosterhoff calculated the mean brightness reported in his Table~1, thus
despite the fact that we have {\it V} magnitudes for V1 and one of the
comparison stars used by Oosterhoff (star $b$ in his chart), we refrain
from attempting to bring the old measurements into a similar magnitude
scale.

{\it V10}. This is a very faint eclipsing binary discovered by
Kaluzny \etal (1997). Its membership in NGC~288 was addressed by Rucinski
(2000) who considers it a cluster member after comparing its value of $M_V$
derived for the cluster distance modulus and the one implied by a
$M_V{-}\log P{-}(B-V)$ relationship.

\begin{figure}[htb] 
\centerline{\includegraphics[width=8cm]{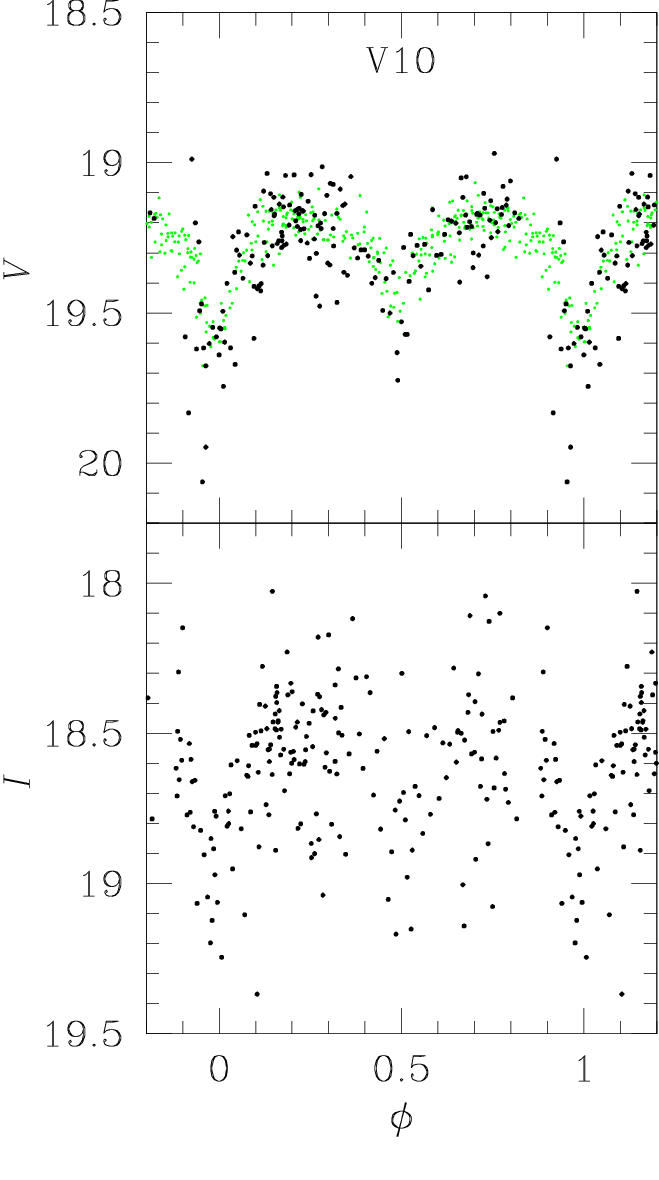}}
\FigCap{{\it V} and {\it I} light curves of V10 with the data from Kaluzny
(1996) and Kaluzny \etal (1997) (green circles) and our data (black
circles). Both data sets have been used to refine the period in Table~4.}
\end{figure}
In Fig.~14 the light curves in our {\it V} and {\it I} data are shown along
with {\it V} data from Kaluzny \etal (1997) phased with the refined period
given in Table~3.

\Section{Summary and Conclusions}
A deep search for variability in the RGB, HB, BS and turnoff point (down to
$V\approx19$~mag) regions of the globular cluster NGC~288 was conducted but
new variables were not found. Thus, we can claim that in the FoV of our
images the census of cluster RR~Lyr stars is complete (except where the
bad pixel column lies, see finding chart in Fig.~6) and that if unknown
SX~Phe stars do exist in the cluster they must be of amplitudes smaller
than the detection limit of our data.

We addressed the apparent off-center distribution of the variables stars.
We investigated the distribution of the SX Phe stars in this cluster by
comparing it to the spatial distribution of all blue straggler stars (BSS).
This is similar to what was done by Kains \etal (2012) for the peculiar
distribution of RR~Lyr stars in NGC~1904 (M79). We first looked at the
angular distribution of the six SX~Phe stars in our sample, \ie how ``close
together'' our SX~Phe stars are, and compared it to randomly drawn samples
of six BSS. We find that the angular distribution of the detected SX~Phe
stars is smaller than 83\% of the randomly drawn samples, which is not
significant given the small sample size. We also looked at the centroid
position of the randomly drawn samples with respect to the center of the
cluster as estimated from our reference image. From this we found that
although visually the stars appear off-center, the offset of their centroid
with the center of the cluster is very close to the most probable value
found from the random samples, with 45\% of the samples having a centroid
further from the center of the cluster than our SX`Phe sample.

The Fourier decomposition of the light curves of one RRab and one RRc star
was performed to calculate the values of [Fe/H], $M_V$, $\log L/\LS$,
$T_{\rm eff}$, the stellar radius and mass, using {\it ad hoc} semi empirical
calibrations (Jurcsik and Kov\'acs 1996, Morgan \etal 2007, Kov\'acs and
Walker (2001) and Kov\'acs (1998).

The mean value of the iron abundance obtained from the two RR~Lyr stars in
the cluster is ${\rm [Fe/H]}_{ZW}=-1.62\pm0.02$ (statistical) $\pm0.14$
(systematic) in the Zinn and West (1984) scale or ${\rm
[Fe/H]}_{UVES}=-1.56\pm0.03$ (statistical) $\pm0.20$ (systematic) in the
scale defined more recently by Carretta \etal (2009). The absolute
magnitudes of the RRab and RRc stars lead to significantly different
distances: from the RRab (V2) we find a distance of 8.99~kpc while from the
RRc we find 7.96~kpc (with internal statistical error of $\approx
0.003$~kpc) but we have noted that the relevant Fourier parameter
$\phi_{21}$ is peculiar in V3. An independent estimate of the cluster
distance was made from the individual stellar distances obtained from the
P--L relation of the SX~Phe stars (Arellano Ferro \etal 2011). We found an
average distance of $8.9\pm0.3$~kpc.

Finally, we identified the two SX~Phe stars, V5 and V9, as double-mode
pulsators. V5 is pulsating in the fundamental and the first overtone while
V9 seems to be pulsating in the first and second overtones plus a
non-radial mode.
 
\Acknow{We are grateful to the TAC of the IAO for generous
telescope time allocation to this project and to the support astronomers of
IAO, at Hanle and CREST (Hosakote) for their very efficient help while
acquiring the data. AAF is gratefull to ESO-Garching for warm hospitallity
during the production of this paper.  This project was supported by
DGAPA-UNAM grant through project IN104612.

We have made a large use of the SIMBAD and ADS services for this
investigation.}

\end{document}